\documentclass[iop]{emulateapj}

\usepackage{braket}

\shorttitle{Viscous Instability \& Layered Accretion}
\shortauthors{Hasegawa \& Takeuchi}

\begin{document}

\title{Viscous Instability Triggered by Layered Accretion in Protoplanetary Disks}

\author{Yasuhiro Hasegawa\altaffilmark{1,2,4} and Taku Takeuchi\altaffilmark{3}}
\affil{$^1$Division of Theoretical Astronomy, National Astronomical Observatory of Japan, Osawa, Mitaka, Tokyo 181-8588, Japan}
\affil{$^2$Jet Propulsion Laboratory, California Institute of Technology, Pasadena, CA 91109, USA}
\affil{$^3$Department of Earth and Planetary Sciences, Tokyo Institute of Technology, Meguro-ku, Tokyo 152-8551, Japan}
\email{yasuhiro@caltech.edu}

\altaffiltext{4}{EACOA fellow}

\begin{abstract}
Layered accretion is one of the inevitable ingredients in protoplanetary disks 
when disk turbulence is excited by magnetorotational instabilities (MRIs).
In the accretion, disk surfaces where MRIs fully operate have a high value of disk accretion rate ($\dot{M}$),
while the disk midplane where MRIs are generally quenched ends up with a low value of $\dot{M}$.
Significant progress on understanding MRIs has recently been made by a number of dedicated MHD simulations, 
which requires improvement of the classical treatment of $\alpha$ in 1D disk models.
To this end, we obtain a new expression of $\alpha$ 
by utilizing an empirical formula that is derived from recent MHD simulations of stratified disks with Ohmic diffusion.
It is interesting that this new formulation can be regarded as a general extension of the classical $\alpha$.
Armed with the new $\alpha$, 
we perform a linear stability analysis of protoplanetary disks that undergo layered accretion,
and find that a viscous instability can occur around the outer edge of dead zones.
Disks become stable in using the classical $\alpha$.
We identify that the difference arises from $\Sigma-$dependence of $\dot{M}$;
whereas $\Sigma$ is uniquely determined for a given value of $\dot{M}$ in the classical approach, 
the new approach leads to $\dot{M}$ that is a multi-valued function of $\Sigma$.
We confirm our finding both by exploring a parameter space as well as by performing the 1D, viscous evolution of disks.
We finally discuss other non-ideal MHD effects that are not included in our analysis, but may affect our results.
\end{abstract}


\section{Introduction} \label{intro}

The origin of disk turbulence is still poorly known \citep[e.g.,][for a recent review]{a11},
while its existence is clearly required to account for the observed disk accretion rate \citep[e.g.,][]{hcg98,wc11}.
One of the most promising ideas is currently the magnetorotational instability (MRI) and the resultant MHD turbulence \citep{bh91a,bh91b}.
The fundamental process of MRIs is angular momentum transport 
that is triggered by magnetic tensions acting on charged particles in differentially rotating disks.
Based on the intensive studies that have been done over the past, more than two decades \citep[e.g.,][for reviews]{bh98,nfg14},
a number of paramount evidence has been obtained that the MRI is a robust instability and can probably operate in protoplanetary disks.
In fact, linear stability analyses and numerical simulations under the ideal MHD limit 
confirm the feasibility of MRIs in disks \citep[e.g.,][]{bh91a,hgb95,dsp10},
while some subtle numerical effects that can affect MRIs still remain to be fully investigated \citep[e.g.,][]{fp07,shb09}.

Substantial efforts have been made to understand MRIs in protoplanetary disks.
This is mainly because gas in the disks is generally dense and cold,
so that the operation of MRIs in disks depends sensitively on non-ideal MHD effects as well as the ionization efficiency of disks.
For the non-ideal MHD effects, Ohmic diffusion is one of the well studied effects \citep[e.g.,][]{j96,sm99,fs03,ll07,oh11}.
It is widely accepted that Ohmic diffusion tends to damp MRIs by smoothing fluctuations in magnetic fields.
The situation becomes more complicated when ambipolar diffusion and Hall term are considered.
For instance, recent numerical simulations show that 
sustainability of MRIs with ambipolar diffusion is determined both by the net vertical magnetic flux and 
by the vertical structure of disks \citep[e.g.,][]{bs11,bs13,sbs13}.
It is important that these results are roughly consistent with classical linear analyses
which simply suggest that the impact of ambipolar diffusion is weak and minor \citep[e.g.,][]{bb94,w99,d04}.
Hall term provides a fundamentally different effect on MRIs,
since it can principally alter the configuration of magnetic fields, rather than diffusing them.
The importance of Hall term is still unclear and currently a matter of debate \citep[e.g.,][]{ss02i,ss02ii,ws12,kl13}.

The ionization efficiency of disks can act as another fundamental parameter for MRIs to be effective in disks \citep[e.g.,][]{smun00,in06}.
This occurs because it regulates coupling between charged particles and magnetic fields threading disks.
In general, disk surfaces and the outer part of disks can efficiently be ionized by UV and X-rays from the central stars \citep[e.g.,][]{ig99,bab07} 
and cosmic rays \citep[e.g.,][]{un09},
while the inner part of disks, where the column density is high enough, has some difficulty to be ionized.
Combining with Ohmic diffusion that becomes dominant in high density regions,
this general trend of MRIs leads to a picture of layered accretion,
wherein MRIs can fully operate in disk surfaces and the outer part of disks,
and the so-called dead zones can be present in the inner part of disks.
This picture was initially proposed by \citet{g96},
and invokes a number of follow-up studies regarding not only disk evolution but also planet formation.
For instance, the presence of dead zones and the resultant layered accretion can affect 
planetesimal formation \citep[e.g.,][]{gnt12,oo13}, planetary migration \citep[e.g.,][]{n05,mpt09,hp11,kl12}, 
and planetary populations \citep[e.g.,][]{il08v,hp13a}.

As discussed above, MHD simulations are clearly powerful tools to investigate in detail the physics of MRIs and the resultant MHD turbulence.
Nonetheless, it still serves as a useful reference to perform the 1D, viscous evolution of disks, 
coupled with the viscous $\alpha-$parameter \citep{ss73}.
This is evident especially when long-term evolution of disks and the effect of MRIs on planet formation are the primary target of investigation.
In this context, interesting questions come up.
Under the circumstances that recent numerical simulations elucidate MRIs more,
how can such an advanced knowledge be incorporated into simple, 1D disk models?
And, how reliable is the classical picture of layered accretion that has widely been used in the community?

To address these questions, we perform a linear stability analysis of protoplanetary disks that undergo layered accretion.
In order to take into account the results of recent numerical simulations and a non-ideal MHD effect, 
we utilize an empirical formula of $\alpha$ 
that is derived from a MHD simulation of stratified disks with Ohmic diffusion and dead zones \citep{oh11}.
We will show below that, while the classical approach always ends with up disks that are stable,
a new approach of layered accretion can trigger a viscous instability around the outer edge of dead zones.
This instability is a consequence of $\alpha$ that becomes a non-monotonic function of the surface density,
and is very likely to be relevant to disk behaviors reported in the previous studies \citep[e.g.,][]{jks11,met13,frd15}.
We also perform time integration of the viscous evolution of disks 
and confirm our finding.
While a more detailed numerical simulations would be needed,
this instability generates a noticeable, transient structure in the surface density profile.
Thus, our results suggest that, even when the standard 1D viscous evolution of disks is considered,
a more realistic receipt of $\alpha$ should be used to take into account MRIs more accurately and 
to mimic the resultant MHD turbulence with non-ideal terms more realistically.

The plan of this paper is as what follows.
In Section \ref{mod}, we briefly discuss our linear stability analysis in which a dispersion relation of viscous instabilities is derived.
We also summarize our assumptions which are needed to simplify mathematical manipulation.
In Section \ref{app}, we apply the relation to both the classical approach of layered accretion and the new one.
We discuss how a new prescription of $\alpha$ can be obtained from the results of recent numerical simulations with Ohmic diffusion.
We present the results of our stability analysis of protoplanetary disks,
and demonstrate that a viscous instability can occur around the outer edge of dead zones.
In Section \ref{discu}, we explore a parameter space to examine whether or not our results are valid for a certain range of model parameters.
We also compute the 1D, viscous evolution of disks with the new receipt of $\alpha$,
and show that a viscous instability can take place for disks with layered accretion.
In addition, we discuss previous work in which similar instabilities are observed.
We finally touch briefly on other non-ideal MHD terms that are not included in our analysis.
Conclusive remarks are presented in Section \ref{conc}.

\section{Stability analysis of disks} \label{mod}

We develop a formulation with which the stability of protoplanetary disks can be discussed.

\subsection{Viscous evolution}

The basic equation for governing the 1D viscous evolution of gas disks is written as \citep[e.g.,][]{p81}
\begin{equation}
 \label{eq:vis_eq}
 \frac{\partial \Sigma}{\partial t} - \frac{3}{r} \frac{\partial}{\partial r} \left[ r^{1/2} \frac{\partial}{\partial r} (r^{1/2} \nu \Sigma) \right] =0,
\end{equation}
where $\nu = \alpha c_s h$ is the kinematic viscosity, $c_s$ is the sound speed, and $h = c_s / \Omega$ is the pressure scale height.
We adopt the famous $\alpha-$prescription to parameterize the strength of disk turbulence \citep{ss73}.

\subsection{Dispersion relation}

We derive a dispersion relation from equation (\ref{eq:vis_eq}).
The relation will serve as a fundamental basis for the following disk stability analysis.

As described in Section \ref{intro}, 
the presence of dead zones in disks and the establishment of layered accretion there are 
determined by the combination of the strength of magnetic fields that thread disks and the ionization efficiency of disks.
The magnetic field strength is poorly constrained, so that it can generally be treated as a free input parameter in MHD simulations.
The ionization efficiency is regulated by the value of $\Sigma$, grain physics in disks,
and disk ionizing sources such as UV and X-rays from the central stars and cosmic rays.
It is ideal to take account of all these quantities and related processes simultaneously.
In this paper, however, we adopt a steady state profile for magnetic fields,
and assume that the value of $\Sigma$ plays a dominant role in determining the ionization efficiency.
This essentially leads to a situation that 
it is sufficient to focus on $\Sigma$ to examine the stability of disks that undergo layered accretion.
We consider this idealized situation, 
in order to shed light on the evolution of $\Sigma$ for disks with layered accretion 
and investigate the resultant effect on disk stability.

Under the assumption,  a dispersion relation can be given as \citep[e.g.,][for a derivation]{p81},
\begin{equation}
 \label{p81_dr}
\frac{\partial }{\partial \Sigma} (\nu \Sigma) <0,
\end{equation}
where $\nu$ is a function of $\Sigma$ alone.
For simplicity, we adopt the locally isothermal assumption hereafter.
Then, we can expand $\Sigma$ and $\alpha$ as 
\begin{equation}
 \Sigma = \Sigma_0 + \Sigma_1,
\end{equation}
and 
\begin{equation}
 \alpha \simeq   \alpha (\Sigma=\Sigma_0) +  \left.  \frac{\partial \alpha}{\partial \Sigma} \right|_{\Sigma=\Sigma_0} \Sigma_1 \equiv  \alpha_0 + \alpha_1,
\end{equation}
where $x_0$ is the unperturbed quantity and $x_1$ is the perturbed one.
As a result, equation (\ref{p81_dr}) can be rewritten as 
\begin{equation}
 \label{eq:nu_0}
 c_s h \left( \alpha_1 \frac{\Sigma_0}{\Sigma_1} + \alpha_0 \right) \equiv \alpha_{sta} c_s h  \equiv \nu_{sta} <0,
\end{equation}
where both $\alpha_{sta}$ and $\nu_{sta}$ are functions of $\Sigma_0$.
Thus, the unstable condition of disks can be determined by a negative diffusion coefficient;
\begin{equation}
\label{eq:alpha_sta}
\alpha_{sta} <0.
\end{equation}
Note that we have performed a linear analysis with the WKB approximation, 
where $\Sigma_1 \equiv \bar{\Sigma} \exp(ikr + \omega t)$ with $kr \gg 1$,
and confirmed the above relation.
In the following sections, we examine $\alpha_{sta}$ 
that is the main quantity to regulate whether or not disks become unstable.
 
\section{Application to disks with layered accretion} \label{app}

We now apply the above unstable condition to disks that undergo layered accretion.
We examine both a classical approach that has widely been adopted in the literature
and a more realistic approach that can be derived from the results of recent numerical simulations.
We will show below that the latter approach of layered accretion can end up with disks that can become unstable. 

\subsection{Disk models} \label{disk_app}

We model disks with layered accretion.
As discussed in Section \ref{mod},
we assume that $\Sigma$ is the most crucial quantity 
to regulate the excitation of layered accretion.
Under the assumption, a simplified formulation of layered accretion is widely used
by introducing two important values of $\Sigma$; 
the one is $\Sigma_{crit}$ that is a critical value of the disk surface density 
that determines whether or not disks undergo layered accretion for a given value of $\Sigma$.
The other is $\Sigma_{AZ}$ that is the surface density of the active region in which MRIs fully operate.
When $\Sigma < \Sigma_{crit}$, the active region is extended from the disk surface to the midplane.
As a result, disks do not experience layered accretion (i.e., $\Sigma_{AZ} \simeq \Sigma$).
On the other hand, layered accretion can be realized when $\Sigma > \Sigma_{cirt}$. 
For this case, dead zones can exist in the disk midplane,
and equivalently $\Sigma_{AZ} \simeq \Sigma_{crit}$.
Thus, we can develop a disk model with layered accretion
by describing $\Sigma_{AZ} / \Sigma $ as
\begin{eqnarray}
 \label{eq:sigma_act}
 \frac{\Sigma_{AZ}}{\Sigma} & \simeq & \frac{\mbox{min}[ \Sigma_{crit}, \Sigma ] }{ \Sigma } = \mbox{min}[ \sigma^{-1}, 1]  \\  \nonumber
                                             & = & ( \mbox{max}[ \sigma, 1] )^{-1}  \approx  (  \sigma^{\gamma} +1 )^{-1/ \gamma}  ,      
\end{eqnarray}
where $\sigma = \Sigma / \Sigma_{crit}$ and $\gamma$ is a parameter to quantify $\mbox{max}[ \sigma, 1] $.
Note that $\Sigma_{AZ} / \Sigma $ (or $\sigma$) will be shown up frequently in the following discussion.
Based on our preliminary study of $\Sigma_{AZ}$ that is calculated from the ionization balance in disks, 
we find that the most likely value is $\gamma \ga 2$,
which will be used in the following sections.

\subsection{Classical approach} \label{class_app}

\begin{figure}
\begin{center}
\includegraphics[width=8cm]{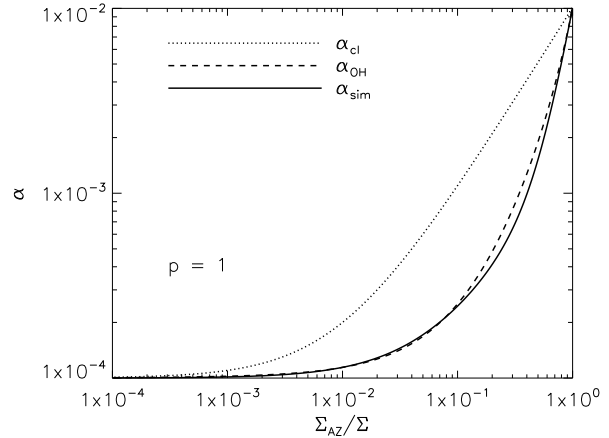}
\caption{The value of $\alpha$ as a function of $\Sigma_{AZ} / \Sigma$.
As an example, we adopt that $\alpha_{AZ} = 10^{-2}$ and $\alpha_{DZ} = 10^{-4}$.
It is assumed that $\beta$ is constant. (Equivalently, $p =1$, see Section \ref{new_alpha_sta}).
We plot $\alpha_{OH}$, using the empirical formula derived by \citet{oh11} (see equation (\ref{eq:alpha_oh})).
Our fitting ($\alpha_{sim}$) reproduces $\alpha_{OH}$ very well (see equation (\ref{eq:alpha_sim2})).
It is obvious that the classical approach ($\alpha_{cl}$) gives the most rapid increase of $\alpha$ than 
both $\alpha_{OH}$ and $\alpha_{sim}$,
which is a consequence that $\dot{M} \propto \Sigma$ for $\alpha_{cl}$.}
\label{fig1}
\end{center}
\end{figure}

We examine a classical treatment of layered accretion.
Assuming that $\alpha_{AD}$ and $\alpha_{DZ}$ is the strength of turbulence in active and dead zones, respectively,
the effective value of $\alpha$ ($\alpha_{cl}$) is generally given as \citep[e.g.,][]{kl07,mpt09,hp11}
\begin{eqnarray}
 \label{eq:alpha_cl}
 \alpha_{cl} & \equiv & \frac{\Sigma_{AZ} \alpha_{AZ} + (\Sigma - \Sigma_{AZ}) \alpha_{DZ} }{\Sigma}  \\ \nonumber
                     & \simeq &  \alpha_{DZ} + \alpha_{AZ} \left( \frac{\Sigma_{AZ}}{\Sigma} \right),
\end{eqnarray}
where it is assumed that both $\alpha_{AZ}$ and $\alpha_{DZ}$ are constant and that $\alpha_{AZ} \gg \alpha_{DZ}$.
Figure \ref{fig1} shows how $\alpha_{cl}$ behaves as a function of $\Sigma_{AZ} / \Sigma$.
It is important to emphasize that this classical picture (equation (\ref{eq:alpha_cl})) explicitly supposes 
that the disk accretion rate ($\dot{M} \propto \alpha \Sigma$) is linearly proportional to $\Sigma$.
In other words, $\Sigma$ is uniquely specified for a given value of $\dot{M}$.

We utilize equation (\ref{eq:sigma_act}) with
an assumption that $\sigma = \sigma_0 + \sigma_1$ (where $|\sigma_1 / \sigma_0| \ll 1$).
Then $\alpha_{cl}$ can be written as 
\begin{equation}
 \alpha_{cl} \simeq \alpha_{cl,0} + \alpha_{cl,1},
\end{equation}
where
\begin{eqnarray}
 \alpha_{cl,0} & \equiv & \alpha_{DZ} + \alpha_{AZ}  (  \sigma_0^{\gamma} +1 )^{-1/ \gamma}, \\
 \alpha_{cl,1} & \equiv & - \alpha_{AZ}  \frac{\sigma_{0}^{\gamma -1} \sigma_1}{ (  \sigma_0^{\gamma} +1 )^{1 + 1/ \gamma} }. 
\end{eqnarray}
Since disk instabilities can be determined by the condition that $\alpha_{sta} < 0$, 
we compute $\alpha_{sta}$, using equation (\ref{eq:nu_0}),
which ends up with
\begin{equation}
\label{eq:alpha_sta_cl}
\alpha_{sta} =  \alpha_{DZ} +  \frac{ \alpha_{AZ}  }{ (  \sigma_0^{\gamma} +1 )^{1 + 1/ \gamma} } >0.
\end{equation}

Thus, we find that disks that undergo layered accretion become always stable 
when the classical approach is adopted (see equation (\ref{eq:alpha_sta})).
As discussed below, however, a new formulation of $\alpha$ 
that is motivated by recent numerical simulations
leads to a different expression of $\alpha$, 
and hence to an alternative outcome on the disk stability analysis.

\subsection{An empirical formula of $\alpha$ by MHD simulations} \label{alpha_emp}

Before introducing our new prescription of $\alpha$,
we summarize an empirical formula that is derived from MHD simulations \citep{oh11}.
We adopt the formula to obtain a new expression of $\alpha$  (see Section \ref{new_alpha}).

\citet{oh11} have recently performed vertically stratified simulations of MHD turbulence,
wherein Ohmic diffusion is included to account for a non-ideal MHD effect and the presence of dead zones.
They have explored the parameter space and obtained an empirical formula for the viscous parameter ($\alpha$) 
that is approximately expressed as a function of three parameters; 
the plasma beta ($\beta_{z}$), the floor value of $\alpha$ ($\alpha_{floor}$), 
and the ratio of the vertical extent of MRI-active zones ($h_{AZ}$) to the disk scale height ($h$). 
The plasma beta that involves the net vertical magnetic flux is defined as
\begin{equation}
 \beta_z \equiv \frac{8 \pi \rho_{mid} c_s^2}{\left\langle {{B_Z}} \right\rangle ^2} 
          = \frac{4 \sqrt(2 \pi) \Sigma c_s \Omega}{\left\langle {{B_Z}} \right\rangle ^2},
\end{equation}
where $\rho_{mid}$ is the gas density at the midplane, 
and $\left\langle {{B_Z}} \right\rangle$ is the averaged vertical strength of magnetic fields. 
The floor value of $\alpha$ ($\alpha_{floor}$) is defined 
such that $\alpha$ becomes the minimum value for fully turbulent disks
(see the horizontal line in the top panel of Figure 2 in \citet{smi10}), 
and is expected to be $\alpha_{floor} \approx 0.01$.

In the end, \citet{oh11} obtain the following form for $\alpha$ (see their equations (28) and (29));
\begin{equation}
\label{eq:alpha_oh}
\alpha_{OH} \simeq \alpha_{floor} \left[  \left( \frac{\beta_z}{\beta_{z0}} \right)^{-1}  + \exp \left( - \delta \frac{h_{AZ}}{h}  \right) \right].
\end{equation}
where $\delta=3.6$ is the numerical factor.
Note that we have simplified the original formula derived by \citet{oh11} in the above expression.
This becomes possible, 
because the first term of their equation (28) is generally negligible,
compared to the first term of their equation (29). 
The normalization factor ($\beta_{z0}$) is defined 
such that $\alpha_{floor}$ becomes a reference value for both the first and the second terms in equation (\ref{eq:alpha_oh}). 
While $\beta_{z0} = (530 / (0.011+0.0043) ) \approx 3.5 \times 10^4$ in \citet{oh11}, 
we here take $\beta_{z0} =10^4$ as a fiducial value.

Equation (\ref{eq:alpha_oh}) contains a number of general features of MHD turbulence triggered by MRIs in protoplanetary disks.
First, the equation is consistent with the numerical results that 
the Maxwell stress is proportional to $\left\langle {{B_Z}} \right\rangle ^2$ \citep[see the first term, e.g.,][]{hgb95}. 
Second, the equation can model a situation that, when disks become fully turbulent and dead zones disappear
(equivalently, the second term becomes unity),
the value of $\alpha$ approaches $\alpha_{floor}$ that is an order of $10^{-2}$. 
This trend is indeed supported by the recent numerical simulations of stratified disks \citep{dsp10,smi10,sab11},
while more simulations would be needed to confirm its robustness.
Thus, the second term of equation (\ref{eq:alpha_oh}) possibly regulates the presence of dead zones, 
given that magnetic fields are weak enough;
when $h_{AZ} \simeq 0$ (i.e., no dead zone), $(\beta_z / \beta_{z0})^{-1} < \exp(- \delta (h_{AZ} / h)) \simeq 1$.
In other words, $\alpha \simeq \alpha_{floor}$.
When $h_{AZ}$ is high, $(\beta_z / \beta_{z0})^{-1} > \exp(- \delta (h_{AZ} / h)) \simeq 0$.
Then $\alpha \simeq \alpha_{floor} (\beta_z / \beta_{z0})^{-1}$.
For this case, dead zones can exist in disks and hence layered accretion can be realized.
In this paper, this switching plays an important role in triggering a viscous instability as discussed below.

\subsection{New prescription of $\alpha$} \label{new_alpha}

We now discuss a new prescritopn of $\alpha$ for disks with layered accretion. 
To proceed, we re-write equation (\ref{eq:alpha_oh}) as a function of $\Sigma_{AZ} / \Sigma$.
Taking account of a definition that $\Sigma_{AZ}/ \ \Sigma = \mbox{erfc} (h_{AZ} / h)$,
we can fit equation (\ref{eq:alpha_oh}) by the following expression ($\alpha_{sim}$);
\begin{equation}
 \label{eq:alpha_sim0}
 \alpha_{sim}   \simeq   \alpha_{floor}  \left[ \left( \frac{\beta_z}{\beta_{z0}} \right)^{-1} + 
                                                             k_1 \left( \frac{\Sigma_{AZ}}{\Sigma} \right) 
                                                           +  k_2 \left( \frac{\Sigma_{AZ}}{\Sigma} \right)^{q}  \right]. 
\end{equation}
Note that we obtain the best fit when $k_1 = 1/7$, $k_2 = 6/7$, and $q =3.6$ (see Figure \ref{fig1}).
In order to make it easy to compare the new approach with the classical one,
we re-define $\alpha_{floor} \equiv \alpha_{AZ}$ and 
\begin{equation}
\alpha_{DZ} \equiv \alpha_{AZ} \frac{\beta_{z0} \left\langle {{B_Z}} \right\rangle ^2_0 }{4 \sqrt(2 \pi) \Sigma_{crit} c_s \Omega },
\end{equation}
where $\left\langle {{B_Z}} \right\rangle _0$ is a characteristic value of $\left\langle {{B_Z}} \right\rangle$.
Then, equation (\ref{eq:alpha_sim0}) becomes
\begin{eqnarray}
 \label{eq:alpha_sim}
 \alpha_{sim}  & \equiv & \alpha_{DZ} 
                                                       \left( \frac{ \left\langle {{B_Z}} \right\rangle }{\left\langle {{B_Z}} \right\rangle_0} \right)^{2} 
                                                       \left( \frac{\Sigma}{\Sigma_{crit}} \right)^{-1}      \\ \nonumber
               & + & \alpha_{AZ} \left[ k_1 \left( \frac{\Sigma_{AZ}}{\Sigma} \right) 
                                                             + k_2 \left( \frac{\Sigma_{AZ}}{\Sigma} \right)^{q}  \right].
\end{eqnarray}
Note that we explicitly label magnetic fields in the above equation to visualize how they relate to the value of $\alpha_{sim}$.

It is interesting that equation (\ref{eq:alpha_sim}) can be regarded as a general extension of equation (\ref{eq:alpha_cl});
the first term of equation (\ref{eq:alpha_sim}) represents 
how the strength of disk turbulence can be affected by magnetic fields threading disks
when dead zones exist there.
In addition, a higher order term of $\Sigma_{AZ} / \Sigma$ appears as the third term in the equation
under the condition that $k_1+k_2 =1$, 
(since $\alpha_{sim} \simeq \alpha_{AZ}$ when $\Sigma_{AZ} \simeq \Sigma$).
This arises because of the exponential function in equation (\ref{eq:alpha_oh}).
Thus, when the results of MHD simulations are taken into account,
the value of $\alpha_{sim}$ increases with $\Sigma_{AZ}$ more gradually than 
what the classical approach expects (see Figure \ref{fig1}, see also equation (\ref{eq:alpha_cl})).
This in turn ends up with $\dot{M}(\propto \alpha_{sim} \Sigma)$ that 
becomes a non-monotonic function of $\Sigma$ (see equation (\ref{eq:alpha_sim})).
We will show below that this trend in $\alpha$ (or $\dot{M}$) significantly changes disk stabilities.

In summary, we adopt the expression of $\alpha$, which can be given as (using equation (\ref{eq:sigma_act}))
\begin{eqnarray}
 \label{eq:alpha_sim2}
 \alpha_{sim} & = & \alpha_{DZ}  \left( \frac{ \left\langle {{B_Z}} \right\rangle }{\left\langle {{B_Z}} \right\rangle_0} \right)^{2} 
                                                       \sigma ^{-1}      \\ \nonumber
                          & + &  \alpha_{AZ} \left[  \frac{1}{7} \left( \sigma^{\gamma} + 1 \right)^{-1 / \gamma} 
                                                       + \frac{6}{7} \left( \sigma^{\gamma} + 1 \right)^{-3.6 / \gamma} \right]. 
\end{eqnarray}

\subsection{Stability analysis} \label{new_alpha_sta}

We are now in a position to perform a stability analysis for $\alpha$ that is given by equation (\ref{eq:alpha_sim2}).

While it has been assumed so far that the magnetic field profile ($ \left\langle {{B_Z}} \right\rangle$) is stationary,
we here attempt to slightly relax this assumption. 
Let us suppose that the evolution of $\Sigma$ can somewhat alter $ \left\langle {{B_Z}} \right\rangle$.
Then, the change in $ \left\langle {{B_Z}} \right\rangle$ ($\Delta \left\langle {{B_Z}} \right\rangle$) can be expressed as a function of $\Sigma$.
For simplicity, we assume that $\Delta \left\langle {{B_Z}} \right\rangle^2 \propto p \Sigma^{p-1}\Delta \Sigma$ 
(i.e., $ \left\langle {{B_Z}} \right\rangle^2 = K(r) \Sigma(r)^p$).
Note that some response of $ \left\langle {{B_Z}} \right\rangle$ to $\Delta \Sigma$ is observed in recent MHD simulations \citep{bs14}.
In this paper, three cases are considered for the value of $p$; $p=0$, $p=1$, and $p=2$.
For the case that $p=0$, it is assumed that magnetic fields ($B_z$) are constant locally,
which corresponds to the case that disk accretion never affects the configuration of magnetic fields that thread disks.
For the case that $p=1$, it is assumed that the plasma $\beta$ is constant locally.
For the case that $p=2$, it is assumed that the ratio between $\Sigma$ and $B_z$ is constant locally,
which is essentially comparable to perfect coupling of magnetic fields with the disk gas.
We adopt this extension, in order to trace the effect of magnetic fields on disk stabilities.

As done in Section \ref{class_app}, we expand $\alpha_{sim}$ as $\alpha_{sim,0} + \alpha_{sim,1}$ with the assumption that 
$\sigma = \sigma_0 + \sigma_1$ (with $|\sigma_1 / \sigma_0| \ll 1$).
Then, $\alpha_{sim,0}$ and $\alpha_{sim,1}$ are given as
\begin{eqnarray}
 \alpha_{sim,0} & \equiv &      \alpha_{DZ}  \sigma_0^{p-1}  \\ \nonumber
                                   &            &  + \alpha_{AZ} \left[  \frac{1}{7} \left( \sigma^{\gamma}_0 + 1 \right)^{-1 / \gamma}
                                                                         +  \frac{6}{7} \left( \sigma^{\gamma}_0 + 1 \right)^{-3.6 / \gamma} \right] , \\
 \alpha_{sim,1} & \equiv &   (p-1) \alpha_{DZ} \sigma_0^{p-2} \sigma_1   \\ \nonumber
                                   &             & -  \alpha_{AZ} \left[  \frac{1}{7} \frac{ 1 }{ \left( \sigma^{\gamma}_0 + 1 \right)^{1 + 1 / \gamma} } 
                                                                 +   \frac{6}{7} \frac{ 3.6 }{ \left( \sigma^{\gamma}_0 + 1 \right)^{1+3.6 / \gamma}}  
                                                           \right]  \sigma_0^{\gamma-1} \sigma_1.
\end{eqnarray}

Consequently, $\alpha_{sta}$ that is important for a stability analysis becomes (see equations (\ref{eq:nu_0}) and (\ref{eq:alpha_sta}))
 \begin{eqnarray}
 \label{eq:alpha_sta_p}
 \alpha_{sta} & \equiv &     p \alpha_{DZ} \sigma_0^{p-1} \\ \nonumber
                          &           &  +  \alpha_{AZ}  \left[   \frac{1}{7} \frac{1}{( \sigma_0^{\gamma} +1)^{1 + 1 / \gamma}}  
                                                                +  \frac{6}{7} \frac{-2.6  \sigma_0^{\gamma} +1}{( \sigma_0^{\gamma} +1)^{1+3.6 / \gamma}}  
                                                     \right] . 
\end{eqnarray}

\begin{figure}
\begin{center}
\includegraphics[width=8cm]{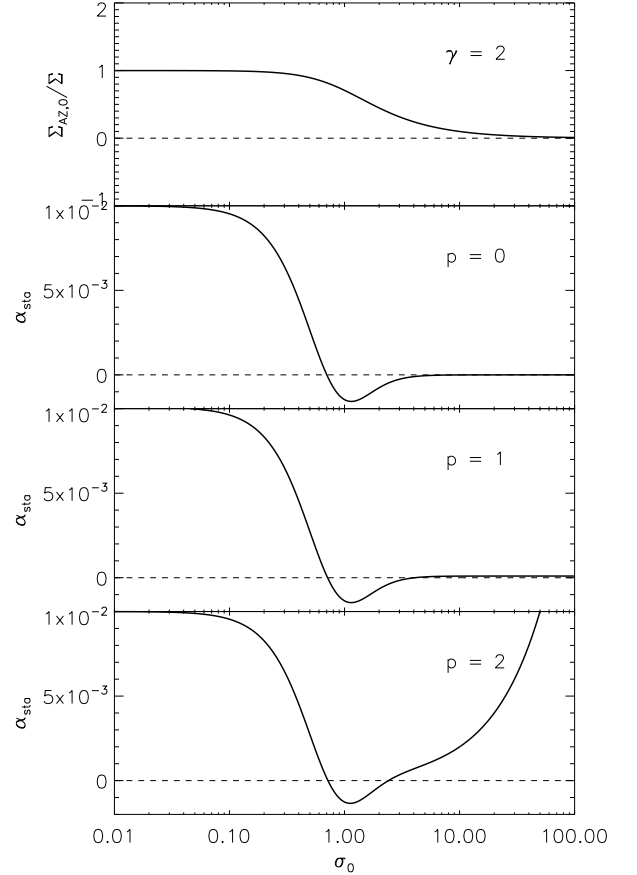}
\caption{The resultant value of $\Sigma_{AZ,0} / \Sigma$ and $\alpha_{sta}$ as a function of $\sigma_0$.
From the top to the bottom, the results of $\Sigma_{AZ,0} / \Sigma$, those of $\alpha_{sta}$ for the case that $p=0$,
those of $\alpha_{sta}$ for the case that $p=1$, those of $\alpha_{sta}$ for the case that $p=2$ are presented, respectively.
The unstable condition ($\alpha_{sta}<0$) is obtained around $\sigma_0 =1$ for all the cases.}
\label{fig2}
\end{center}
\end{figure}

Figure \ref{fig2} shows the resultant behavior of $\alpha_{sta}$.
The case that $\gamma =2$ is considered (see Section \ref{disk_app}).
We plot $\Sigma_{AZ,0} / \Sigma$ on the top,
to illustrate how the parameter $\gamma$ affects $\Sigma_{AZ,0} / \Sigma$ as a function of $\sigma_0$ (see equation (\ref{eq:sigma_act})).
On the second to the bottom, we plot $\alpha_{sta}$ as a function of $\sigma_0$ with $p$ varying;
the second is for the case that $p=0$,
the third is for the case that $p=1$,
and the bottom is for the case that $p=2$.
In our formulation, 
the ratio, $\alpha_{AZ} / \alpha_{DZ}$, will affect the outcome of disk stability analysis.
We have here adopted that $\alpha_{AZ} =10^{-2} $ and $\alpha_{DZ} = 10^{-4} $,
and perform a parameter study about them later (see Figure \ref{fig3}).

One immediately finds that the unstable condition ($\alpha_{sta} <0$) is achieved around $\sigma_0 =1$ for all the cases (see Figure \ref{fig2}).
Our results therefore indicate that disks that undergo layered accretion can be unstable, especially in regions where $\Sigma_{AZ} \la \Sigma$.
The difference with the classical approach arises largely from 
a slower increase of $\alpha$ as a function of $\Sigma_{AZ} / \Sigma$ (see Figure \ref{fig1}),
which finally provides a negative contribution to $\alpha_{sta}$ (see equation (\ref{eq:alpha_sta_p})).
The trend originates from the term $(\Sigma_{AZ} /  \Sigma)^q$ with $q>1$ in equation (\ref{eq:alpha_sim}), 
which involves the results of MHD simulations (see the exponential function in equation (\ref{eq:alpha_oh})).

\section{Discussion} \label{discu}

In the above sections, 
we demonstrate that the new prescription of $\alpha$ based on the recent numerical simulations 
can trigger a viscous instability when disks undergo layered accretion.
Here, we explore a parameter space to examine how sensitive the above results are to model parameters such as $\alpha_{DZ}$.
In addition, we compute 1D viscous evolution of disks and investigate a consequence of the instability for disk structures on a long timescale.
Since similar instabilities have been reported in previous studies,
we also discuss them.
Finally, we comment briefly on other non-ideal MHD terms that have been excluded in our analysis.

\subsection{Parameter study}

We perform a parameter study to investigate how valid are our results 
that are derived from the case that $\alpha_{AZ}=10^{-2}$ and $\alpha_{DZ}=10^{-4}$.
Since all the results can be scaled by $\alpha_{AZ}$, 
we simply change the value of $\alpha_{DZ}$.

\begin{figure*}
\begin{minipage}{17cm}
\begin{center}
\includegraphics[width=8cm]{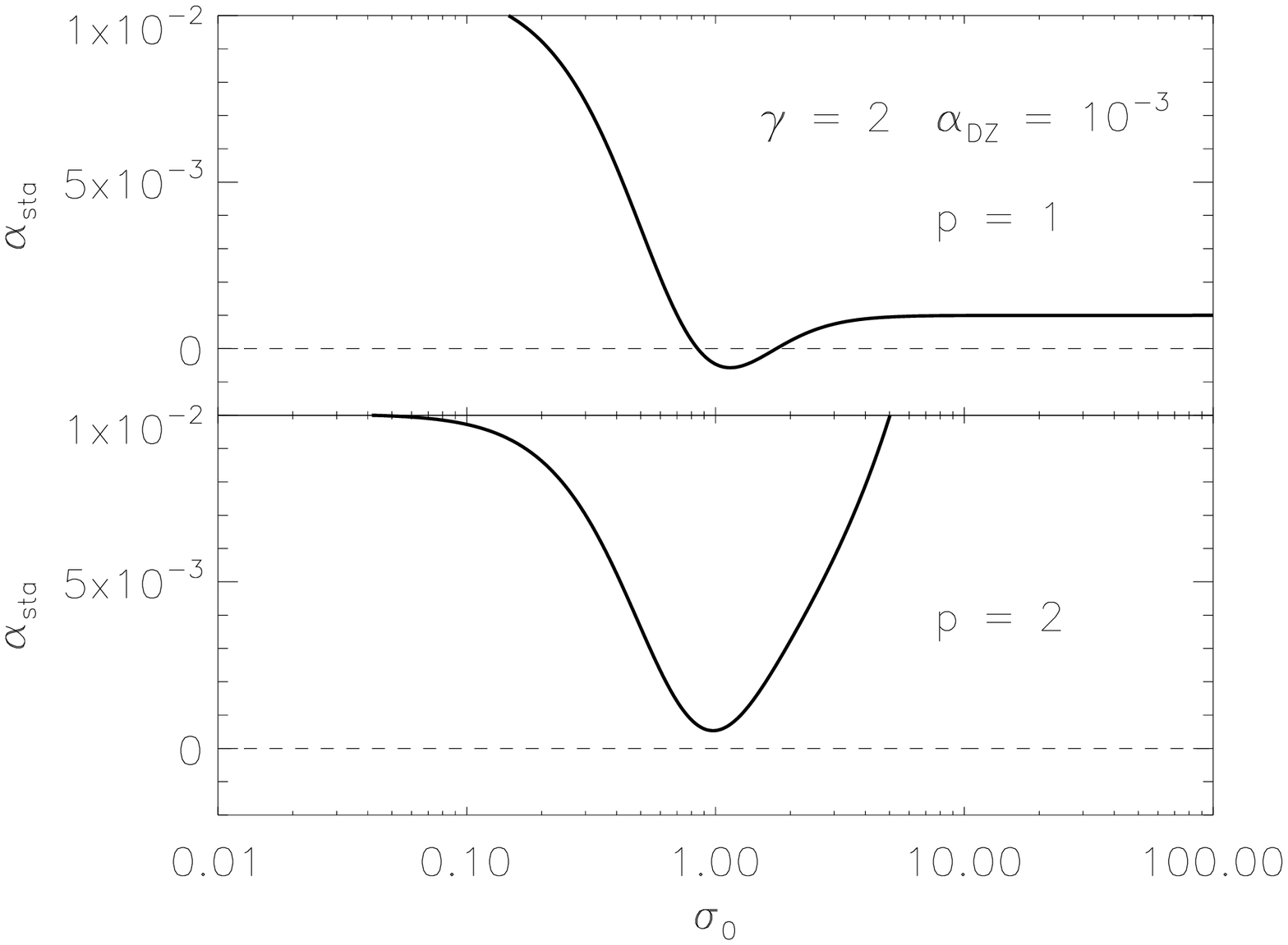}
\includegraphics[width=8cm]{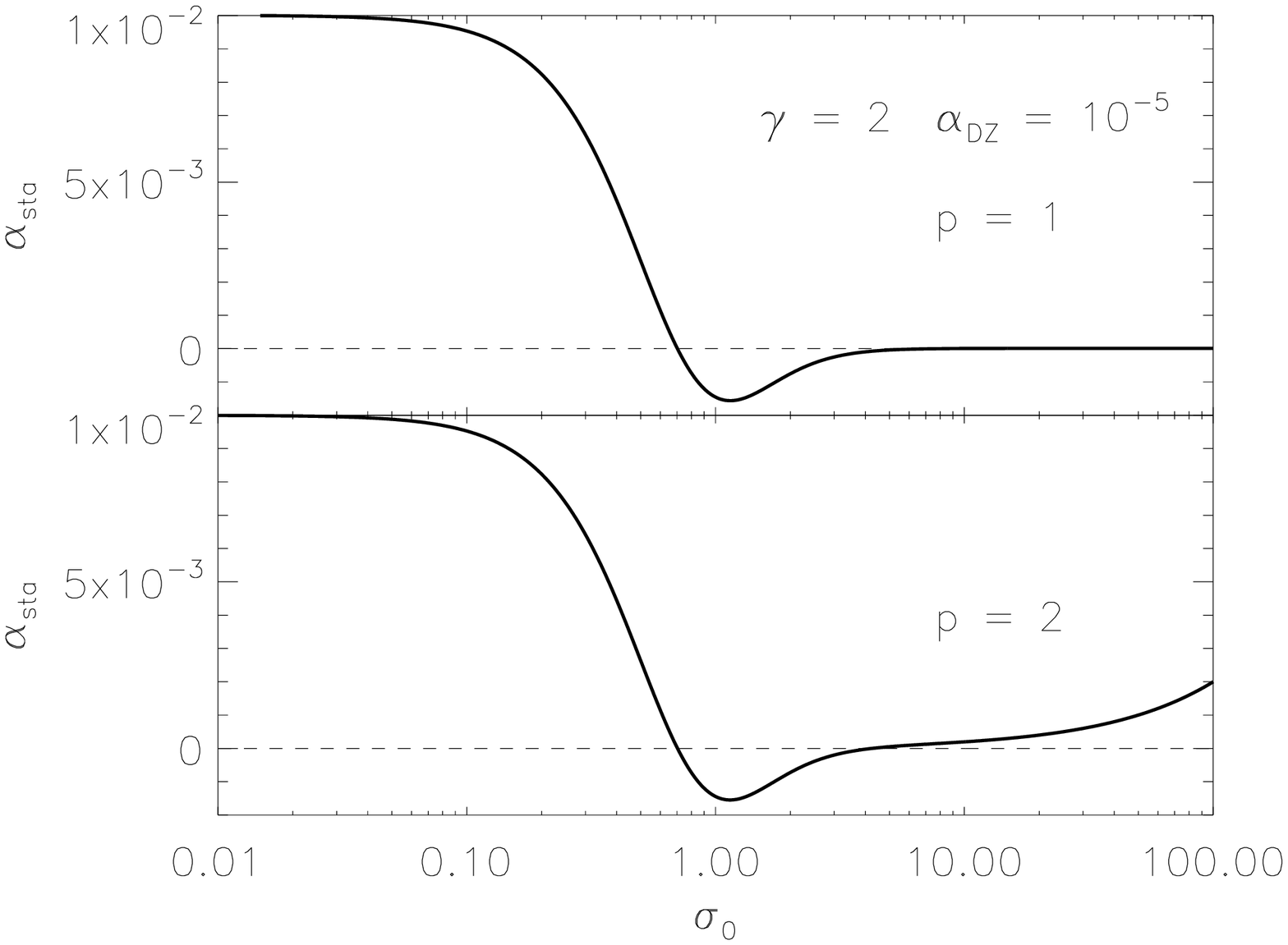}
\caption{Parameter study of $\alpha_{sta}$. 
The left panel shows the results for the case that  $\alpha_{DZ} =10^{-3}$, 
while the right panel is for the case that  $\alpha_{DZ} =10^{-5}$.
We confirm that disks with layered accretion can generally satisfy the unstable condition ($\alpha_{sta}<0$) 
as long as $\alpha_{DZ}/\alpha_{AZ} < 10^{-1}$ and $p<2$ with the usage of the new formulation of $\alpha$ (see equation (\ref{eq:alpha_sim2})).}
\label{fig3}
\end{center}
\end{minipage}
\end{figure*}

Figure \ref{fig3} summarizes its outcome; 
the results for the case that $\alpha_{DZ} =10^{-3}$ are on the left panel,
and those for the case that $\alpha_{DZ} =10^{-5}$ are on the right panel.
Since the dependence of $\alpha_{DZ}$ disappears for the case that $p=0$,
we do not show the corresponding results on each panel.
The results show that, expect for the case that $\alpha_{DZ}=10^{-3}$ and $p=2$,
layered accretion can trigger a viscous instability in disks.
This can be expected, because the term, $p \alpha_{DZ} \sigma_0^{p-1}$, in equation (\ref{eq:alpha_sta_p}) 
provides a positive contribution to $\alpha_{sta}$.
As a result, when a high value of $\alpha_{DZ}$ (equivalently, a high value of $\alpha_{DZ} / \alpha_{AZ}$) is adopted,
disks with layered accretion can be stabilized.

In summary, disks that undergo layered accretion can become unstable
as long as $\alpha_{DZ} / \alpha_{AZ} < 10^{-1}$ and $p <2$ (see equation (\ref{eq:alpha_sim2})).

\subsection{Long-term evolution}

We have so far analyzed a viscous instability based on a linear stability analysis.
We now discuss what is a consequence of the instability on disk evolution.
To this end, we perform time integration of equation (\ref{eq:vis_eq}) and obtain long-term evolution of disk structures.
In the computation, we adopt the viscosity model given by equation (\ref{eq:alpha_sim2}) 
and assume that $\alpha_{AZ} / \alpha_{DZ} = 100$.

\begin{figure}
\begin{center}
\includegraphics[width=8cm]{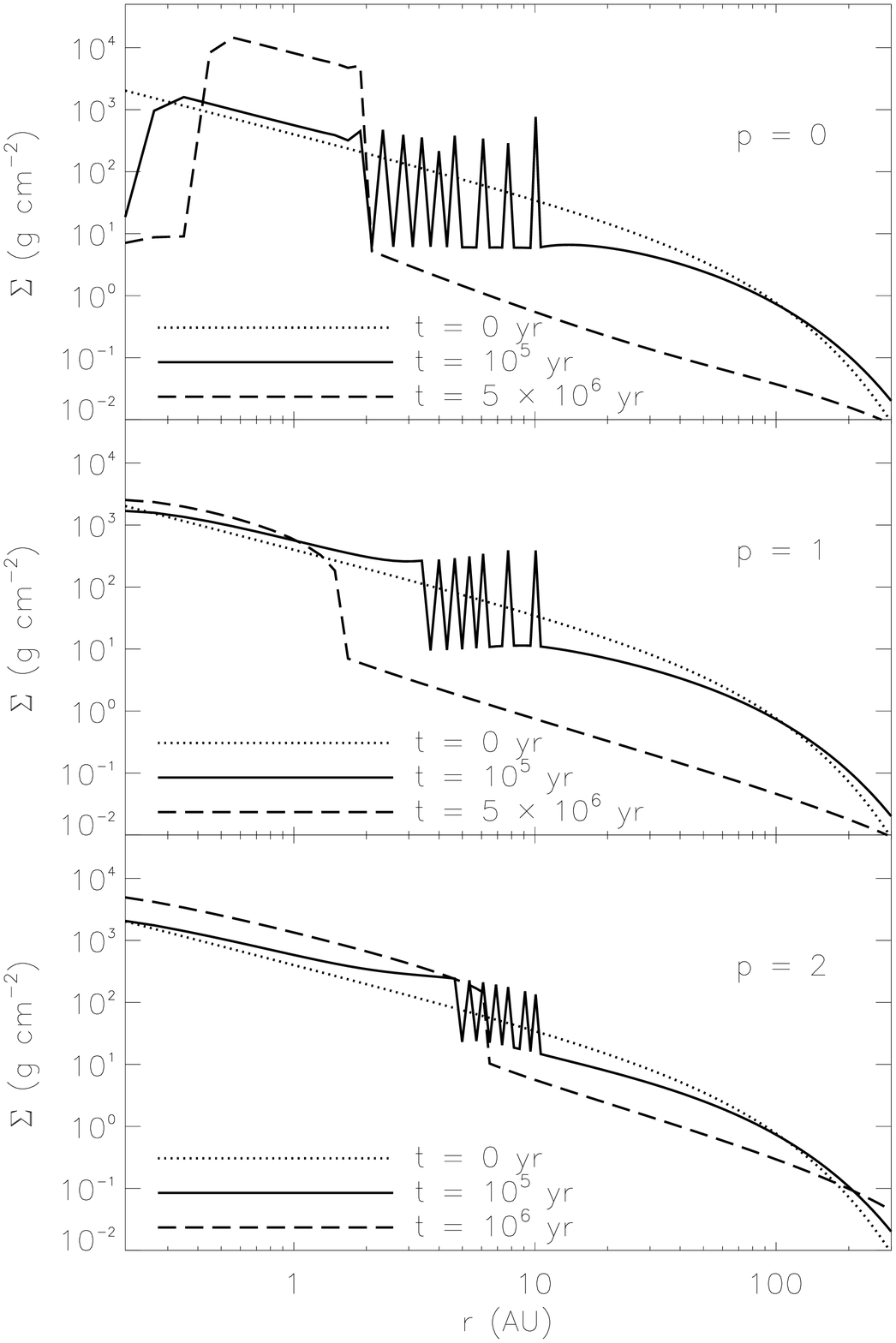}
\caption{Long term evolution of $\Sigma$ for disks that undergo layered accretion.
In the computations, we set that $\alpha_{AZ} =10^{-2}$ and $\alpha_{DZ} = 10^{-4}$.
For the initial condition (see the dotted line), 
$\Sigma$ is given by the similarity solution \citep{lbp74},
while the magnetic field distribution is set so that $\beta_{z}=10^6$ everywhere.
By adjusting the critical surface density ($\Sigma_{crit}$),
the outer edge of dead zones is initially located at $r \simeq 10$ AU.
From the top to the bottom, the results for the case that $p=0$, $p=1$, and $p=2$ are shown, respectively.
The numerical simulations confirm that a viscous instability occurs around the outer edge of dead zones,
which ends up with a noticeable saw-tooth-like structure there.
This result therefore supports our linear analysis.
The saw-tooth-like structure disappears when the disk accretion rate becomes low at $t \simeq 10^6$ yr.}
\label{fig4}
\end{center}
\end{figure}

Figure \ref{fig4} shows how a viscous instability evolves at the outer boundary of dead zones.
The results for the case that $p=0$ are presented on the top panel, 
those for the case that $p=1$ are on the middle panel,
and those for the case that $p=2$ are on the bottom panel. 
We find that the instability grows at $t \lesssim 10^5$ yr for all the cases of $p$,  
and the surface density profile eventually exhibits a prominent saw-tooth-like structure at the outer dead zone edge,
which is located at $r \simeq 10$ AU in our computations.
We notice that the surface density difference between crests and troughs corresponds roughly to the ratio $\alpha_{\rm AZ} / \alpha_{\rm DZ} = 100$,
and that the distance between two crests (or two troughs) is almost identical to the grid interval of the numerical calculation.
We thus consider that the structure does not directly relate to any physical quantity. 
The actual structure would involve physical lengths such as the disk scale height and a characteristic length of thermal diffusion. 
As time goes on, the disk gas accumulates in the inner part of disks,
where dead zones are present.
It is interesting that the saw-tooth-like structure disappears around $t =10^6$ yr.
This suggests that the structure is a transient phenomenon and 
can be maintained only when the mass accretion from the outer disk is high enough (see Figure \ref{fig4}).

The results also show that, for the case that $p=0$, 
the surface density drops locally in the inner part of disks at $5 \times 10^6$ yr (see the top).
This occurs due to the mass accretion rate ($\dot{M} \propto \alpha_{sim} \Sigma$).
For the cases that $p=1$ and $p=2$, 
$\dot{M}$ becomes lower as $\Sigma$ decreases (see equation (\ref{eq:alpha_sim})).
As a result, $\Sigma$ reduces with time monotonically at all the positions.
For the case that $p=0$, contrastingly,
$\dot{M}$ becomes constant in dead zones even if $\Sigma$ decreases with time.
This leads to a more efficient mass accretion onto the central star especially in dead zones, 
and hence generates a local decrement in $\Sigma$ there.
It is interesting that such a rapid decrease arises when $B_z$ is stationary.
We will discuss the feature more in a subsequent paper.

\subsection{Relevance to previous studies}

As discussed above, the new prescription of $\alpha$ formulates the mass accretion rate 
that becomes a non-monotonic function of $\Sigma$ (see equation (\ref{eq:alpha_sim})).
We have found that such a formulation eventually triggers a viscous instability (see Figure \ref{fig2}).
This is entirely different from the classical approach that was initially proposed by \citet{g96}.
In the approach, the mass accretion rate is linearly proportional to $\Sigma$ (see equation (\ref{eq:alpha_cl})),
and hence disks become stable against a viscous instablity.

While this is the first attempt to perform a linear stability analysis in the framework of the $\alpha-$parameter to examine a viscous instability,
excitation of similar instabilities has already been discussed; 
for instance, \citet{jks11} have undertaken a pioneering work of a viscous instability in layered accretion,
wherein local linear stability analysis is developed in a shearing sheet formalism.
As we have found in this paper,
they also confirm that viscous instability can occur in disks with layered accretion.
As another example, 
\citet{frd15} have performed global, 3D MHD simulations of stratified disks with Ohmic diffusion,
and investigated the structure of outer dead zone edges.
They have found that a density gap followed by a density bump is formed there,
which is very likely to result from a viscous instability discussed in this paper \citep[also see][]{met13}.
We will undertake a more comprehensive study and examine what is the relationship between our and their work.

Another instability that involves layered accretion may be a gravo-magneto disk instability \citep[e.g.,][]{alp01,zhg10,mrl14}. 
The fundamental origin of this instability is mass pile-up in dead zones;
when the mass accretion rate beyond dead zones is higher than the value of $\dot{M}$ in the layered accretion region,
then the disk mass can be piled up there.
This accumulation enhances the viscous heating and heats up the surrounding disk gas. 
Once the disk temperature reaches a critical value 
at which disks can be thermally ionized,
MRIs become fully active even in dead zones, 
and outburst events arise.
Since this instability generates a limit cycle that can be characterized by the "S-curve" in the $\nu \Sigma- \Sigma$ diagram, 
it can be viewed as an analogue of the classical thermal instability \citep[e.g.,][]{bl94,lc04}.
In general, this instability requires a high value of $\Sigma$ and can occur at the early disk evolution stage.
On the other hand, our viscous instability is very likely to be excited in the middle and/or late stage of disk evolution.  
In fact, the formula derived by \citet{oh11} is valid for $\beta_z \la 10^6$ at which $\Sigma$ would be too low to thermally ionize the disk gas.
Thus, it is unlikely that these two instabilities would overlap.
Nonetheless, it may be interesting to examine whether or not the resultant disk structure caused by our viscous instability can 
serve as an initial condition for the gravo-magneto instability;
the local peaks in $\Sigma$ may be high enough to trigger the gravo-magneto instability (see Figure \ref{fig4}),
while the gas pressure and/or other disk instabilities may prevent the instability from occurring.
We will leave a more detailed study for future work.

We simply emphasize here that 
a key point for the viscous instability we have found is that,
when dead zones are present in disks,
the mass accretion rate is entirely independent of $\Sigma$ \citep{oh11,gnt12};
assuming that vertical magnetic flux is weak enough, 
$\alpha \propto 1/\beta_z$ (see Section \ref{alpha_emp}).
This in turn indicates that $\Sigma$ can easily fluctuates even when the constant mass accretion rate is realized. 
In our formulation, the exponential function in equation (\ref{eq:alpha_oh}) serves as a perturbation to trigger a viscous instability.
This property is completely missed in the classical $\alpha$-viscosity model. 
The fluctuation in $\Sigma$ can modify the ionization rate in disks,
which would eventually change a dominant non-ideal MHD term there. 
In this paper, we demonstrate that a small perturbation in $\Sigma$ alters effectiveness of Ohmic diffusion 
and results in a viscous instability. 

\subsection{Other non-ideal MHD terms}

We have adopted the expression of $\alpha$ based on \citet{oh11} that takes into account only Ohmic diffusion (see equation (\ref{eq:alpha_oh})).
This putative model has been used, 
in order to qualitatively investigate how non-ideal MHD effects can affect disk evolution and what is their consequence on disk structures.

What happens if both ambipolar diffusion and Hall effect would be included in our analysis?
It has been suggested over two decades that these two terms play an important role in disk evolution \citep{bb94,w99}.
Nonetheless, it has recently been made it possible to perform detailed numerical simulations \citep[][]{b14,b15,lkf14,slk15}.
These simulations show that these terms possibly alter the qualitative behavior of mass accretion in protoplanetary disks significantly.
While it is obvious that more dedicated work would be needed 
to elucidate the quantitative behavior of disk evolution under the action of both ambipolar diffusion and Hall effect ,
it is worth expanding our investigation to examine whether or not similar instability can take place in protoplanetary disks
when these non-ideal MHD effects are included.

\section{Conclusions} \label{conc}

MRIs are one of the important processes that can be considered as an origin of disk turbulence.
The sustainability of MRIs in protoplanetary disks is still under debate, 
because disks provide cool and dense environments for gas, 
and hence both non-ideal MHD terms and the ionization efficiency of disks play a crucial role in exciting MRIs.
This in turn invokes a proposal of layered accretion
in which disk accretion rates can be composed of two layers;
in disk surface layers, MRIs are fully effective and a high value of accretion rate can be achieved.
For the midplane of disks, contrastingly, ionization fraction can drop rapidly due to shielding of high energy photons 
by a high value of column density.
Coupling with Ohmic diffusion, such a low degree of ionization fraction can generate the so-called dead zone there,
and end up with a low value of disk accretion rate.

Ever since the pioneering work of layered accretion by \citet{g96},
significant progress has been made on understanding of MRIs in protoplanetary disks and the resultant MHD turbulence.
It is considerably important and quite timely to incorporate such a progress that is driven largely by detailed MHD numerical simulations 
into a classical picture of layered accretion.

We have undertaken a linear stability analysis of protoplanetary disks that undergo layered accretion.
In our analysis, we have examined both the classical approach and a new one 
that is motivated by the results of recent numerical simulations.
We have demonstrated that, when the classical picture of layered accretion is adopted (see equation (\ref{eq:alpha_cl})),
disks become always stable (see equation (\ref{eq:alpha_sta_cl})).
This is a simple reflection of disk accretion rates ($\dot{M}$) that are monotonic functions of $\Sigma$.

We have obtained a new prescription of $\alpha$, 
by making use of the results of recent MHD simulations \citep{oh11}.
In the simulations, Ohmic diffusion is included to model the presence of dead zones in disks.
Following their empirical formula (see equation (\ref{eq:alpha_oh})),
we have come up with a new recipe of $\alpha$  (see equation (\ref{eq:alpha_sim})).
We then have performed a linear stability analysis, 
and demonstrated that such a new expression of $\alpha$ can generate a viscous instability around the outer edge of dead zones (see Figure \ref{fig2}).
This occurs, because the resultant $\dot{M}$ becomes a non-monotonic function of $\Sigma$ 
(see the exponential function in equation (\ref{eq:alpha_oh}) and the third term in equation (\ref{eq:alpha_sim})).
More physically, this involves a general feature of MRIs that, when dead zones are present,
$\dot{M}$ is regulated largely by magnetic fields, not by $\Sigma$ (since $\alpha \propto 1 / \beta_z$, see Section \ref{alpha_emp}).
As a result, $\Sigma$ can vary considerably even if $\dot{M}$ becomes constant. 
This trend is entirely neglected in the classical view of layered accretion.
We have therefore shown a viscous instability triggered by layered accretion in protoplanetary disks,
by adopting a more realistic formulation of $\alpha$.

We have also performed a parameter study to examine how valid our findings are.
We have shown that a viscous instability can take place for a wide range of model parameters (see Figure \ref{fig3}).
To investigate the effect of the instability on disk structures,
we have computed time evolution of $\Sigma$ (see equation (\ref{eq:vis_eq}))
under the action of $\alpha-$viscosity given by equation (\ref{eq:alpha_sim}).
We have demonstrated that a prominent, saw-tooth-like structure appears around the outer edge of dead zones (see Figure \ref{fig4}).
While a more careful examination would be required for this transient phenomena,
the numerical results can be viewed as favored evidence for the viscous instability found by our linear analysis.
It is also interesting that previous studies have observed a density gap and bump around the outer edge of dead zones,
which may be generated by similar instabilities.
A more comprehensive study is desired to fully understand viscous instabilities and the resultant effect on disk structure.
We have briefly discussed other non-ideal MHD effects that have been excluded in our analysis.
It is obvious that more intensive studies would be needed to elucidate their effects on our results.

Finally, we would like to stress the importance of global magnetic field configuration. 
The mass accretion rate is mainly determined by the magnetic flux as discussed in this paper. 
Here we treated the magnetic flux (and consequently $\alpha_{DZ}$) as free input parameters, 
but in reality these would be determined by magnetic flux transport in disks \citep{lpp94}. 
The magnetic flux transport should be intensively explored to determine actual $\alpha_{DZ}$ 
\citep[e.g.,][for recent progress on the flux transport in protoplanetary disks]{otm14,g014,to14}.
We will undertake it in a subsequent paper.


\acknowledgments

The authors thank Mario Flock, Satoshi Okuzumi, and Neal Turner for stimulating discussions,
and an anonymous referee for useful comments on our manuscript.
The part of this research was carried out at the Jet Propulsion Laboratory, California Institute of Technology, 
under a contract with the National Aeronautics and Space Administration.
Y.H. is currently supported by Jet Propulsion Laboratory, California Institute of Technology,
and has previously been by EACOA Fellowship that is supported by East Asia Core Observatories Association which consists of 
the Academia Sinica Institute of Astronomy and Astrophysics, the National Astronomical Observatory of Japan, the National Astronomical 
Observatory of China, and the Korea Astronomy and Space Science Institute.  
T.T. is supported by Grants-in-Aid for Scientific Research, Nos. 23103005, 26103704, and 26400224 from MEXT of Japan.






\bibliographystyle{apj}          

\bibliography{apj-jour,adsbibliography}    

\begin{thebibliography}{63}
\expandafter\ifx\csname natexlab\endcsname\relax\def\natexlab#1{#1}\fi

\bibitem[{Armitage(2011)}]{a11}
Armitage, P.~J. 2011, \araa, 49, 195

\bibitem[{{Armitage} {et~al.}(2001){Armitage}, {Livio}, \& {Pringle}}]{alp01}
{Armitage}, P.~J., {Livio}, M., \& {Pringle}, J.~E. 2001, \mnras, 324, 705

\bibitem[{{Bai}(2014)}]{b14}
{Bai}, X.-N. 2014, \apj, 791, 137

\bibitem[{{Bai}(2015)}]{b15}
---. 2015, \apj, 798, 84

\bibitem[{{Bai} \& {Stone}(2011)}]{bs11}
{Bai}, X.-N. \& {Stone}, J.~M. 2011, \apj, 736, 144

\bibitem[{{Bai} \& {Stone}(2013)}]{bs13}
---. 2013, \apj, 769, 76

\bibitem[{{Bai} \& {Stone}(2014)}]{bs14}
---. 2014, \apj, 796, 31

\bibitem[{{Balbus} \& {Hawley}(1991)}]{bh91a}
{Balbus}, S.~A. \& {Hawley}, J.~F. 1991, \apj, 376, 214

\bibitem[{{Balbus} \& {Hawley}(1998)}]{bh98}
---. 1998, Reviews of Modern Physics, 70, 1

\bibitem[{Bell \& Lin(1994)}]{bl94}
Bell, K.~R. \& Lin, D. N.~C. 1994, \apj, 427, 987

\bibitem[{{Bergin} {et~al.}(2007){Bergin}, {Aikawa}, {Blake}, \& {van
  Dishoeck}}]{bab07}
{Bergin}, E.~A., {Aikawa}, Y., {Blake}, G.~A., \& {van Dishoeck}, E.~F. 2007,
  Protostars and Planets V, 751

\bibitem[{{Blaes} \& {Balbus}(1994)}]{bb94}
{Blaes}, O.~M. \& {Balbus}, S.~A. 1994, \apj, 421, 163

\bibitem[{{Davis} {et~al.}(2010){Davis}, {Stone}, \& {Pessah}}]{dsp10}
{Davis}, S.~W., {Stone}, J.~M., \& {Pessah}, M.~E. 2010, \apj, 713, 52

\bibitem[{{Desch}(2004)}]{d04}
{Desch}, S.~J. 2004, \apj, 608, 509

\bibitem[{{Fleming} \& {Stone}(2003)}]{fs03}
{Fleming}, T. \& {Stone}, J.~M. 2003, \apj, 585, 908

\bibitem[{{Flock} {et~al.}(2015){Flock}, {Ruge}, {Dzyurkevich}, {Henning},
  {Klahr}, \& {Wolf}}]{frd15}
{Flock}, M., {Ruge}, J.~P., {Dzyurkevich}, N., {Henning}, T., {Klahr}, H., \&
  {Wolf}, S. 2015, \aap, 574, A68

\bibitem[{{Fromang} \& {Papaloizou}(2007)}]{fp07}
{Fromang}, S. \& {Papaloizou}, J. 2007, \aap, 476, 1113

\bibitem[{Gammie(1996)}]{g96}
Gammie, C.~F. 1996, \apj, 457, 355

\bibitem[{{Gressel} {et~al.}(2012){Gressel}, {Nelson}, \& {Turner}}]{gnt12}
{Gressel}, O., {Nelson}, R.~P., \& {Turner}, N.~J. 2012, \mnras, 422, 1140

\bibitem[{{Guilet} \& {Ogilvie}(2014)}]{g014}
{Guilet}, J. \& {Ogilvie}, G.~I. 2014, \mnras, 441, 852

\bibitem[{Hartmann {et~al.}(1998)Hartmann, Calvet, Gullbring, \&
  D{'}Alessio}]{hcg98}
Hartmann, L., Calvet, N., Gullbring, E., \& D{'}Alessio, P. 1998, \apj, 495,
  385

\bibitem[{{Hasegawa} \& {Pudritz}(2011)}]{hp11}
{Hasegawa}, Y. \& {Pudritz}, R.~E. 2011, \mnras, 417, 1236

\bibitem[{{Hasegawa} \& {Pudritz}(2013)}]{hp13a}
---. 2013, \apj, 778, 78

\bibitem[{{Hawley} \& {Balbus}(1991)}]{bh91b}
{Hawley}, J.~F. \& {Balbus}, S.~A. 1991, \apj, 376, 223

\bibitem[{{Hawley} {et~al.}(1995){Hawley}, {Gammie}, \& {Balbus}}]{hgb95}
{Hawley}, J.~F., {Gammie}, C.~F., \& {Balbus}, S.~A. 1995, \apj, 440, 742

\bibitem[{Ida \& Lin(2008)}]{il08v}
Ida, S. \& Lin, D. N.~C. 2008, \apj, 685, 584

\bibitem[{{Igea} \& {Glassgold}(1999)}]{ig99}
{Igea}, J. \& {Glassgold}, A.~E. 1999, \apj, 518, 848

\bibitem[{{Ilgner} \& {Nelson}(2006)}]{in06}
{Ilgner}, M. \& {Nelson}, R.~P. 2006, \aap, 445, 205

\bibitem[{{Jin}(1996)}]{j96}
{Jin}, L. 1996, \apj, 457, 798

\bibitem[{{Johansen} {et~al.}(2011){Johansen}, {Kato}, \& {Sano}}]{jks11}
{Johansen}, A., {Kato}, M., \& {Sano}, T. 2011, in IAU Symposium, Vol. 274, IAU
  Symposium, ed. A.~{Bonanno}, E.~{de Gouveia Dal Pino}, \& A.~G. {Kosovichev},
  50--55

\bibitem[{Kretke \& Lin(2007)}]{kl07}
Kretke, K.~A. \& Lin, D. N.~C. 2007, \apjl, 664, L55

\bibitem[{Kretke \& Lin(2012)}]{kl12}
---. 2012, \apj, 755, 74

\bibitem[{{Kunz} \& {Lesur}(2013)}]{kl13}
{Kunz}, M.~W. \& {Lesur}, G. 2013, \mnras, 434, 2295

\bibitem[{{Lesur} {et~al.}(2014){Lesur}, {Kunz}, \& {Fromang}}]{lkf14}
{Lesur}, G., {Kunz}, M.~W., \& {Fromang}, S. 2014, \aap, 566, A56

\bibitem[{{Lesur} \& {Longaretti}(2007)}]{ll07}
{Lesur}, G. \& {Longaretti}, P.-Y. 2007, \mnras, 378, 1471

\bibitem[{{Lodato} \& {Clarke}(2004)}]{lc04}
{Lodato}, G. \& {Clarke}, C.~J. 2004, \mnras, 353, 841

\bibitem[{{Lubow} {et~al.}(1994){Lubow}, {Papaloizou}, \& {Pringle}}]{lpp94}
{Lubow}, S.~H., {Papaloizou}, J.~C.~B., \& {Pringle}, J.~E. 1994, \mnras, 267,
  235

\bibitem[{{Lynden-Bell} \& Pringle(1974)}]{lbp74}
{Lynden-Bell}, D. \& Pringle, J.~E. 1974, \mnras, 168, 603

\bibitem[{{Martin} \& {Lubow}(2014)}]{mrl14}
{Martin}, R.~G. \& {Lubow}, S.~H. 2014, \mnras, 437, 682

\bibitem[{Matsumura {et~al.}(2009)Matsumura, Pudritz, \& Thommes}]{mpt09}
Matsumura, S., Pudritz, R.~E., \& Thommes, E.~W. 2009, \apj, 691, 1764

\bibitem[{{Mohanty} {et~al.}(2013){Mohanty}, {Ercolano}, \& {Turner}}]{met13}
{Mohanty}, S., {Ercolano}, B., \& {Turner}, N.~J. 2013, \apj, 764, 65

\bibitem[{Nelson(2005)}]{n05}
Nelson, R.~P. 2005, \aap, 443, 1067

\bibitem[{{Okuzumi} \& {Hirose}(2011)}]{oh11}
{Okuzumi}, S. \& {Hirose}, S. 2011, \apj, 742, 65

\bibitem[{{Okuzumi} {et~al.}(2014){Okuzumi}, {Takeuchi}, \& {Muto}}]{otm14}
{Okuzumi}, S., {Takeuchi}, T., \& {Muto}, T. 2014, \apj, 785, 127

\bibitem[{{Ormel} \& {Okuzumi}(2013)}]{oo13}
{Ormel}, C.~W. \& {Okuzumi}, S. 2013, \apj, 771, 44

\bibitem[{Pringle(1981)}]{p81}
Pringle, J.~E. 1981, \araa, 19, 137

\bibitem[{Sano {et~al.}(2000)Sano, Miyama, Umebayashi, \& Nakano}]{smun00}
Sano, T., Miyama, S., Umebayashi, T., \& Nakano, T. 2000, \apj, 543, 486

\bibitem[{{Sano} \& {Miyama}(1999)}]{sm99}
{Sano}, T. \& {Miyama}, S.~M. 1999, \apj, 515, 776

\bibitem[{{Sano} \& {Stone}(2002{\natexlab{a}})}]{ss02i}
{Sano}, T. \& {Stone}, J.~M. 2002{\natexlab{a}}, \apj, 570, 314

\bibitem[{{Sano} \& {Stone}(2002{\natexlab{b}})}]{ss02ii}
---. 2002{\natexlab{b}}, \apj, 577, 534

\bibitem[{Shakura \& Sunyaev(1973)}]{ss73}
Shakura, N.~I. \& Sunyaev, R.~A. 1973, \aap, 24, 337

\bibitem[{{Simon} {et~al.}(2011){Simon}, {Armitage}, \& {Beckwith}}]{sab11}
{Simon}, J.~B., {Armitage}, P.~J., \& {Beckwith}, K. 2011, \apj, 743, 17

\bibitem[{{Simon} {et~al.}(2013){Simon}, {Bai}, {Stone}, {Armitage}, \&
  {Beckwith}}]{sbs13}
{Simon}, J.~B., {Bai}, X.-N., {Stone}, J.~M., {Armitage}, P.~J., \& {Beckwith},
  K. 2013, \apj, 764, 66

\bibitem[{{Simon} {et~al.}(2009){Simon}, {Hawley}, \& {Beckwith}}]{shb09}
{Simon}, J.~B., {Hawley}, J.~F., \& {Beckwith}, K. 2009, \apj, 690, 974

\bibitem[{{Simon} {et~al.}(2015){Simon}, {Lesur}, {Kunz}, \&
  {Armitage}}]{slk15}
{Simon}, J.~B., {Lesur}, G., {Kunz}, M.~W., \& {Armitage}, P.~J. 2015, ArXiv
  e-prints

\bibitem[{{Suzuki} {et~al.}(2010){Suzuki}, {Muto}, \& {Inutsuka}}]{smi10}
{Suzuki}, T.~K., {Muto}, T., \& {Inutsuka}, S.-i. 2010, \apj, 718, 1289

\bibitem[{{Takeuchi} \& {Okuzumi}(2014)}]{to14}
{Takeuchi}, T. \& {Okuzumi}, S. 2014, \apj, 797, 132

\bibitem[{{Turner} {et~al.}(2014){Turner}, {Fromang}, {Gammie}, {Klahr},
  {Lesur}, {Wardle}, \& {Bai}}]{nfg14}
{Turner}, N.~J., {Fromang}, S., {Gammie}, C., {Klahr}, H., {Lesur}, G.,
  {Wardle}, M., \& {Bai}, X.-N. 2014, Protostars and Planets VI, 411

\bibitem[{{Umebayashi} \& {Nakano}(2009)}]{un09}
{Umebayashi}, T. \& {Nakano}, T. 2009, \apj, 690, 69

\bibitem[{{Wardle}(1999)}]{w99}
{Wardle}, M. 1999, \mnras, 307, 849

\bibitem[{{Wardle} \& {Salmeron}(2012)}]{ws12}
{Wardle}, M. \& {Salmeron}, R. 2012, \mnras, 422, 2737

\bibitem[{Williams \& Cieza(2011)}]{wc11}
Williams, J.~P. \& Cieza, L.~A. 2011, \araa, 49, 67

\bibitem[{Zhu {et~al.}(2010)Zhu, Hartmann, \& Gammie}]{zhg10}
Zhu, Z., Hartmann, L., \& Gammie, C. 2010, \apj, 713, 1143

\end{thebibliography}

\end{document}